# Depletion Gilding: An Ancient Method for Surface Enrichment of Gold Alloys

Amelia Carolina Sparavigna

1 – Department of Applied Science and Technology, Politecnico di Torino, Torino, Italy

**ABSTRACT.** Ancient objects made of noble metal alloys, that is, gold with copper and/or silver, can show the phenomenon of surface enrichment. This phenomenon is regarding the composition of the surface, which has a percentage of gold higher than that of the bulk. This enrichment is obtained by a depletion of the other elements of the alloy, which are, in some manner, removed. This depletion gilding process was used by pre-Columbian populations for their "tumbaga", a gold-copper alloy, to give it the luster of gold.

**Introduction:** A phenomenon often encountered when ancient objects made of noble metal alloys are analyzed is that of their surface enrichment. Let us consider, for instance, a statuette made of an alloy of gold with copper and/or silver; it can occur that its surface has a percentage of gold higher than that of the bulk. This enrichment can be due to an addition of gold on the surface, or to a depletion process during which the less chemically stable elements leach out causing the surface composition to change. In both cases, the local percentage of gold is increased consequently. Therefore, both processes are gilding processes, the second being known as "depletion gilding". It happens because a specific depletion process had been applied to the surface of the object or because it had been buried for a rather long time [1] (of course, besides such a slow gilding process, time is causing a long series of damaging and corrosive effects [2]). Masters of the depletion gilding were the pre-Columbian populations of America that used it for their "tumbaga", an alloy of gold and copper, to give the luster of gold to the objects made of it. In this paper, we will discuss some aspects of tumbaga and depletion gilding.

**Gilding:** The term "gilding" covers several techniques for applying a gold leaf or a gold powder to solid surfaces, in order to have a thin coating of this metal on objects. Several methods of gilding exist, including hand application, chemical gilding and electroplating. These are additive methods, which act by depositing gold onto the surface of objects usually made of a less precious material. Among the techniques of gilding, some are quite old. Fire-gilding of metals for instance goes back at least to the 4th century BC, and was known to Pliny the Elder and Vitruvius. Fire-gilding is a process by which an amalgam of gold is applied to metallic surfaces. Objects are set on fire and mercury volatilizes, leaving a film of gold or a gold-rich amalgam on the surface. About the coating with gold leaf, Pliny is also telling the following. "When copper has to be gilded, a coat of quicksilver is laid beneath the gold leaf, which it retains in its place with the greatest tenacity: in cases, however, where the leaf is single, or very thin, the presence of the quicksilver is detected by the paleness of the colour" [3].

As previously told, besides the gilding obtained by the abovementioned techniques, we have also the subtractive process of depletion gilding. In this gilding, some material is removed to increase the purity of the gold. Of course, gold must be already present on the surface of the object. For this reason, this gilding procedure can be applied only to objects composed by gold alloys, usually gold with copper and/or silver. The gilding is performed by removing the metals, which are not gold. These metals are etched away from the surface by means of the use of some acids or salts, often combined with the action of heat. Of course, there is no gold addition, because the object already contains gold.

**Depletion gilding**: Depletion gilding is based on the property of gold of being resistant to oxidation or corrosion by most chemicals, whereas many other metals, such as copper and silver, are not so. Therefore an object, cast for instance by an alloy of gold with copper and silver, can be immersed in a suitable acid or packed in a salt, which attacks the copper and silver in the object's surface. The action of acid or salt is transforming these elements to some copper/silver compounds that can be removed from the object's surface by washing or heating, or by using a brick dust [4, 5]. The result is a thin layer of nearly pure gold on the surface of the object. Often it is necessary to repeat this procedure several times, making the resulting surface soft and spongy with a dull appearance. For this reason, most depletion-gilded objects are burnished to make their surfaces more durable and give them a more attractive polished finish.

Depletion gilding was widely used in antiquity. A historical and technical introduction of this gold surface enrichment is given in [6], which discusses how goldsmiths have used the depletion gilding technique for "coloring the gold". The process requires some skill to execute it properly, but it is technologically simple. Moreover, it is requiring materials that were available to most ancient civilizations, those that were able of making alloys.

For what concerns the color of gold, let us note that pure gold is slightly reddish yellow in color. Other colors can be produced making alloys with silver, copper, nickel and zinc in various proportions, producing white, yellow, green and red golds [7] (see Appendix for some data). In the case of an alloy of gold and copper, the result is a red or yellow-red color. These alloys were used especially in the pre-Columbian Meso- and South America. Known as "tumbaga", this material was used widely both for castings and for hammered metal works. A further depletion gilding was giving to these objects the color and luster of pure gold.

Today, gold alloys have many applications in dentistry, jewelry and industrial areas too (let us note that gold is used for corrosion protection of electrically conductive surfaces [8]). For economic reasons then, much effort has been made to lower the gold content in the bulk of such alloys. As a consequence, the surface enrichment of low gold alloys became an interesting subject of researches [9]: in this reference, we find the modern methods for the creation of a gold-enriched surface on a gold alloy by depletion processes. Experiments tell that the additions of sodium chloride to pure water speed up the oxidation of copper to copper chloride, which is dissolved at the metal-solution interface. Additions of sodium sulfide to pure water should also speed up the oxidation of copper to cuprous or cupric sulfide, but these compounds are insoluble in water, and then they are tarnishing the alloy [9]. Let us remember that, in pure water, copper dissolves regardless of being in a gold alloy or as metal [9].

**American goldsmiths**: However, how did the pre-Columbian populations of Meso- and South America a depletion gilding? This question was the subject of several researches made by archaeologists. In [4], we can find a description of techniques. It is also told that was Gonzalo Fernandez de Oviedo (1478-1557), to give a hint on pre-Columbian depletion gilding, writing that the pre-Columbian goldsmiths knew how to use a certain herb for gilding objects made of debased gold. The alloys used were generally of two types [4]. One type is composed by the tumbaga copper-gold alloys produced with differing gold contents, the other was that of pale greenish-white ternary silver-gold-copper alloys, containing a high proportion of silver, similar to the Mediterranean electrum and widely used in Peru.

For tumbaga, a depletion gilding technique was the following: the object made of tumbaga was rubbed with the juice of a plant and then heated so that it assumed a gold coloration. This process was repeated many times to improve the colour and increase the superficial gold contain. It is believed that the plant was a species of oxalis and that the juice contained oxalic acid [4]. For objects made of alloys of electrum type, they were probably gilded using a cementation process or by using some aqueous pastes [4]. In the first process, the object was placed in a crucible and surrounded with a powdered mixture containing alum, common salt and brick dust [4]. The crucible and its contents were heated. The mixture reacted with the surface of the object, forming chlorides of silver, copper and other impurity metals. The chlorides were absorbed by the brick dust [4]. Probably, additional ingredients may have been used too. After cooling, and subsequently washing the object, the brilliance of the surface was increased by burnishing [4]. The second method was that of immersing the object in an aqueous paste or solution of alum, iron sulfate and salt at room temperature. After about ten days [4], the object was washed in a strong salt solution and then heated to convert the spongy, gold-enriched surface to a smooth and compact surface. Both cementation and aqueous methods work equally well on electrum and tumbaga [4].

**Phase diagram of gold and copper**: Tumbaga is an alloy of gold and copper then. However, we can tell more about these two metals together. A study of the phase diagram of copper and gold shows that they are completely soluble in each other with eutectic type low melting point, occurring at a composition of 80.1% gold at 911 °C. In the Figure 1, the phase diagram is shown, adapted from [10]. Phase diagrams of gold with other elements, such as platinum, silver, nickel and cobalt are given in Ref.11. Let us remember that a naturally occurring alloy of gold and silver exists, the electrum.

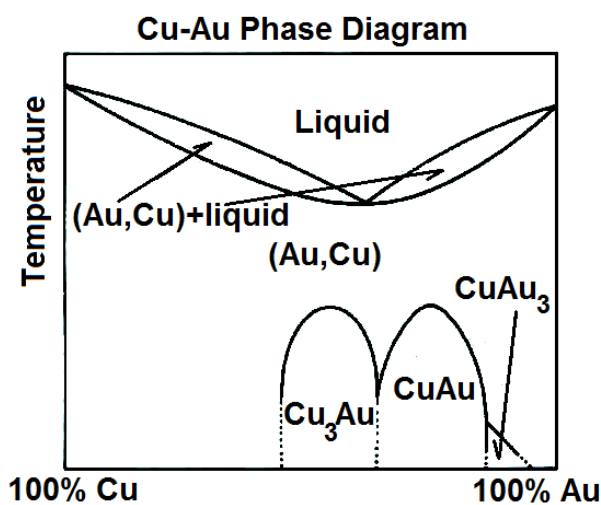

*Fig. 1. Phase diagram Au/Cu*

In the Figure 1, it is easy to see the eutectic composition. The word "eutectic" comes from the Greek word "eutektos", that is, "easily melted". At the eutectic-composition, an alloy of two or more metals, when heated to its melting point, completely changes from solid to liquid at the same temperature [12]. Thus, the eutectic-composition is characterized by being the first alloy-composition to melt during heating [12], such as the last to freeze during cooling.

The rounded shapes at the bottom of the diagram in Fig.1 show the regions where ordered phases exist. According to [10], these ordered phases are usually harder than the disordered alloy of the same composition, and they may make the process of working and annealing to shape more difficult. Moreover, the quenched alloys between about 85% gold and 50% gold are softer than the alloys that are allowed to cool slowly in air (quenching is the rapid cooling of a work piece). This is the opposite of what happens in alloys such as iron and carbon, where the material is hardened by quenching because of the formation of martensitic phase [10]. For the gold-copper alloys, the softening by quenching process happens because it is suppressing the formation of the ordered phases, which need some time to form. As told in [10], South American populations used water quenching after annealing in order to make their alloys easier to work to shape and to avoid embrittlement.

**The raft of El Dorado**: The alloys of gold with copper or silver were produced by the pre-Columbian people to create wonderful statues and ornaments. In the Figure 2, it is shown one of such objects. It is the Muisca Raft, obtained in a lost-wax casting by the Muisca culture in a region which currently corresponds to the center of Colombia. A recent study on Muisca metallurgy shows that gold alloys were especially composed for votive metalwork [13], and in fact, the Raft is a votive object. Today, it is exhibited at the Gold Museum in Bogota. The Raft refers to the ceremony of El Dorado, during which Muisca chief, after covering his body with gold powder, dove into the Guatavita Lake. Then, El Dorado was El Hombre Dorado, the Golden Man.

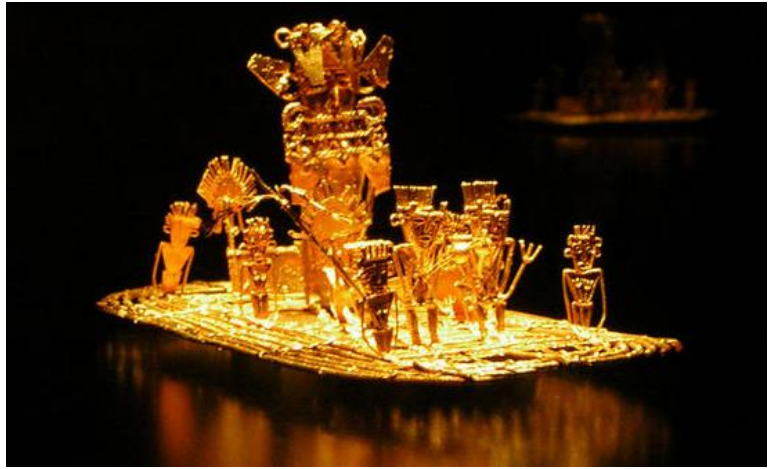

*Fig. 2. The Muisca Raft (Courtesy: Wikipedia). It is a representation of the legend of El Dorado. The cacique at the center of the raft is surrounded by attendants and oarsmen*

The legends surrounding El Dorado changed over time, so that it became a golden city or a lost kingdom full of gold. Many expeditions were made in the search for El Dorado: among the most famous there was that led by Sir Walter Raleigh [14, 15]. All the expeditions did not find El Dorado but mapped a large part of South America.

After failing in discovering El Dorado and its gold mines, the Spaniard conquistadores that had promised their king a mass of gold in return for investing in the transatlantic voyage, resorted to looting the treasuries of the local chiefs and the grave goods of cemeteries [16]. However, as Shakespeare writes in his play "The Merchant of Venice", "all that glitters is not gold": when Spaniard soldiers began to melt down the mass of the glittering ornaments, they discovered that they had not pure gold, but an alloy debased with large amounts of copper. The result was that a large part of beautiful objects, such as those held in the Gold Museum, were plundered by the Spaniards and melted into "tumbaga" bars for transport across the Atlantic. Hernan Cortes and his men for instance improvised a manufacture of such metallic bars [17].

Because all the metals that reached Europe were melted back into their constituent metals in Spain, there is only an example of such a load, a group of over 200 tumbaga bars, discovered in the remains of an unidentified shipwreck (around 1528), off Grand Bahama Island. This shipwreck was found in 1993 [17]. Since we have told of Hernan Cortes, the conquistador who caused the fall of Aztec Empire and brought Mexico under the rule of the King of Castile, let us mention an interesting article on the metallurgy of Aztecs [18]. In it, is mentioned the pioneering research work of Dora M.K. de Grinberg and others on the metallurgical skills of the pre-Columbian population [19]. De Grinberg, an Argentinian archeologist working in Mexico, uncovered ample evidence that the ancient American metalworkers were far more skilled than had previously been supposed [18].

**Tumbaga:** The word "tumbaga" is not native to any language of the area of Meso- or South America. But it is not a Spanish word too. It is coming from Malay and means copper [20]. The historical documentation on ancient American gold alloys begins with Columbus, who reported that the word "guanin" was employed to express these alloys. Washington Irving, in his "Life of Columbus", wrote that in 1503 Columbus was on the Mosquito Coast. "There was no pure gold to be met with here, all their ornaments were of guanine; but the natives assured the Adelantado that in proceeding along the coast, the ships would soon arrive at a country where gold was in abundance" [21]. In the reports of Columbus, it is evident his quest for gold. The same happens for other explorers. In a 1546 communication to his king, Juan Perez de Tolosa reported on a population of the Northwestern Venezuela that, in addition to possessing gold and other precious metals, had ornaments of a copper-gold alloy called "carcuri". Similar reports appear in the writing of Pedro de Cieza de Leon, who explored the Cuca Valley of Northern Colombia during 1532-1550.

In a book of 1760 [22], written by Antonio de Ulloa (1716-1795), Spanish general and explorer, we find other information about gold: "In the district of Choco are many mines of Lavadero, or wash gold … There are also some, where mercury must be used, the gold being enveloped in other metallic bodies, stones and bitumens. Several of the mines have been abandoned on account of the platina; a substance of such resistance, that, when struck on an anvil of steel, it is not easy to separate … In some of these mines the gold is found mixed with the metal called tumbaga, or copper, and equal to that of the east". Antonio de Ulloa uses the word "tumbaga" for copper then. He continues telling that "its most remarkable quality is that it produces no verdigrease (verdigris), nor is corroded by any acids, as common copper is well known to be" [22]. In fact, if we treat tumbaga with an acid, copper is dissolved off the surface. On the surface, it remains a shiny layer of nearly pure gold. As previously discussed, the use of an acid produces a process of depletion gilding, not the verdigris. Note that, in the description made by Antonio de Ulloa, there is also mentioned another material, the platina, that is, the platinum.

Among the first modern reports about tumbaga, there is that by G. Créqui-Montfort and P. Rivet, published in 1919 [23], who described the tumbaga in Colombia. The documentation of a similar pre-Columbian alloy with depletion gilding to produce a golden surface is given in the Ref.24. In a report of 1949, W. Root is comparing the physical properties of tumbaga with those of unalloyed gold and copper. In his review [25], Root tells that tumbaga seems to have originated in Colombia or Venezuela before AD 1000 and spread to Ecuador and Peru. But, in a discussion about gilding [26], Heather Lechtman et al. tell that the depletion gilding was first developed by the Moche culture of Peru, about AD 100-800. Therefore, depletion methods of gilding used in Peru, from this center of origin, spread north into Ecuador, Columbia, Venezuela, Panama till Mexico [26].

**In nanotechnologies**: Today, tumbaga has an important role in nanotechnology. In fact, the gold-copper alloys are emerging as an important catalyst. In [27], the authors investigated the phase diagrams of various polyhedral nanoparticles, made of gold-copper alloy. In these particles, the researchers revealed a gold enrichment at the surface, like in tumbaga, leading to a kind of core-shell structure, analogous to the surface enrichment of archaeological artifacts. The most stable structures of the nanoparticles were determined to be the dodecahedron, truncated octahedron, and icosahedron with a Cu-rich core/Au-rich surface. In [28], nanorods of $AuCu_3$ had been investigates, in particular to determine the catalytic activity of them, when different surface ligands are used.

Besides nanoparticles, in catalysis, sensing, and other areas, porous gold is used [29-32]. This material is made by dealloying gold alloys [29]. In fact, this relatively new material is like the surface of tumbaga, the spongy gold which is produces by dealloying the surface layer with gilding depletion. For what concerns porous gold, let us conclude with an interesting feature of the layer of gold on tumbaga, discussed by Stuart J. Fleming in Ref.16. Fleming tells that tumbaga has a "self-healing" property. When corrosion happens, some gold atoms are set free by it. These atoms can migrate and seal the minute channels, which are originated by the corrosive attack. For this reason, objects made of a relatively gold-rich tumbaga can retain for a long time their original luster. This property of self-healing of gold alloys could be interesting for nanotechnologies too, where surfaces have a relevant role, due to the reduced dimensions of involved materials.

**Appendix on colors of gold-copper alloys**: Here a table on the color and chemical composition of some alloys with karat number 18 [7]. Let us remember that 24 kt gold is pure gold. The designations 18 kt, 14 kt, or 10 kt indicate how much pure gold is present in the mix: 18 kt gold (75% gold) has 18 parts gold and 6 parts of another metal(s), 14 kt gold (58.3% gold) has 14 parts gold and 10 parts of another metal(s), and so on for 12 kt and 10 kt gold. 10 kt gold is the minimum karat designation that can still be called gold in the US [7].

*Table 1*

| Color of Gold | Alloy Compositions Containing Copper |
|---|---|
| Yellow Gold (22 kt) | Gold 91.67%, Silver 5%, Copper 2%, Zinc 1,33% |
| Red Gold (18 kt) | Gold 75%, Copper 25% |
| Rose Gold (18 kt) | Gold 75%, Copper 22.25% Silver 2.75% |
| Pink Gold (18 kt) | Gold 75%, Copper 20%, Silver 5% |
| Gray-White Gold (18 kt) | Gold 75%, Iron 17%, Copper 8% |
| Light Green Gold (18 kt) | Gold 75%, Copper 23%, Cadmium 2% |
| Green Gold (18 kt) | Gold 75%, Silver 20%, Copper 5% |
| Deep Green Gold (18 kt) | Gold 75%, Silver 15%, Copper 6%, Cadmium 4% |